\documentclass[english,12pt]{article}
\usepackage{amssymb}
\usepackage{amsmath}
\usepackage{babel}
\usepackage{array}
\usepackage{amssymb}
\usepackage{pstricks}
\usepackage{pst-node}
\date{}
\title{The next stage: quantum game theory}
\author{Edward W. Piotrowski\\ Institute of Theoretical Physics,
University of Bia\l ystok,\\ Lipowa 41, Pl 15424 Bia\l ystok,
Poland\\ e-mail: ep@alpha.uwb.edu.pl \\
 Jan S\l adkowski\\ Institute of Physics, University of Silesia, \\ Uniwersytecka
4, Pl 40007 Katowice, Poland \\ e-mail: sladk@us.edu.pl }
\begin{document}
\def\meter{\mbox{$\frown\hspace{-.9em}{\lower-.4ex\hbox{$_\nearrow$}}$}}
\def\circs{\mbox{$\hspace{-.3em}\bigcirc\hspace{-.69em}{\lower-.35ex\hbox{$_{\mathsf S}$}}$}}
\def\circh{\mbox{$\hspace{-.16em}\bigcirc\hspace{-.76em}{\lower-.35ex\hbox{$_{\mathsf H}$}}$}}
\maketitle
\begin{abstract}
\noindent Recent development in quantum computation and
quantum information theory  allows to extend the scope of game
theory for the quantum world. The paper presents the history,
basic ideas and recent development in quantum game theory. On
grounds of the discussed material, we reason about possible future
development of quantum game theory and its impact on information
processing and  the emerging information society.
 \end{abstract}
 \vspace{5mm}

\section{Introduction}

The emerging of global information infrastructure caused one of
the main paradigm shifts in human history: information is becoming
a crucial if not the most important resource. Recently the
scientific community has became more and more aware that
information processing is a physical phenomenon and that
information theory is inseparable from both applied and
fundamental physics.  Attention to the very physical aspects of
information processing revealed new perspectives of computation,
cryptography and  communication methods.  In most of the cases
quantum description of the system provides advantages over the
classical situation. Game theory, the study of (rational) decision
making in conflict situations, seems to ask for a quantum version.
For example, games against nature \cite{mil} include those for
which nature is quantum mechanical. Does quantum theory offer more
subtle ways of playing games? Game theory considers strategies
that are probabilistic mixtures of pure strategies. Why cannot
they be intertwined in a more complicated way, for example
interfered or entangled? Are there situations in which quantum
theory can enlarge the set of possible strategies? Can quantum
strategies be more successful than classical ones? And if the
answer is yes are they of any practical value? John von~Neumann is
one of the founders of both game theory \cite{NM} and quantum
theory, is that a meaningful coincidence? In this paper we would
like to convince the reader that the research on  quantum game
theory cannot be neglected because present technological
development suggest that sooner or later someone would take full
advantage of quantum theory and may use quantum strategies to beat
us at some game. Cryptography and communication methods seem to be
the more probable battle fields but who can be sure? The paper is
organized as follows. We will begin by presenting a detailed
analysis of a simple example given by David A\mbox{.} Meyer \cite{Mey} that
will illustrate the general idea of a quantum game and methods of
gaining an advantage over "classical opponent". Then we will
attempt at giving a definition of a quantum game and review
problems that have already been discussed in the literature.
Finally we will try to show some problems that should be addressed
in the near future. In the following discussion we will use
quantum theory as a language but the broadcasted message would be
that it can be used as a weapon.

\section{Quantum game prehistory and history}

It is not easy to give the precise date of birth of quantum game
theory. Quantum games have probably been camouflaged since the
very beginning of the  quantum era because a lot of experiments
can be reformulated in terms of game theory. Quantum game theory
began with works of Wiesner on quantum money \cite{Wie}, Vaidman,
who probably first used the term  game in quantum context
\cite{Vai}, and  Meyer \cite{Mey}  and Eisert et al \cite{Eis} who
first formulated their problems in game theory formalism. Possible
applications of quantum games in biology were discussed by Iqbal
and Toor \cite{IT1}, in economics by Piotrowski and S\l adkowski
\cite{PS1,PS2}. Flitney and Abbott quantized Parrondo's paradox
\cite{Abb}. David Meyer put forward a fabulous argument for
research on quantum game theory that we are going to retell here
\cite{Mey}. He describes a game that is likely to be played by two
characters of the popular TV series {\it Star Trek: The Next
Generation\/}, Captain Picard and Q. Suppose they play the modern
version of the penny flip game that is implemented as a {\it
spin--flip game}\/ (there probably are no coins on a starship).
Picard is to set an electron in the spin up state, whereupon they
will take turns (Q, then Picard, then Q) flipping the spin or not,
without being able to see it. Q wins if the spin is  up when they
measure the electron's state. This is a two--person {\it zero--sum
strategic game}\/ which might be analyzed using the payoff matrix:
$$
\vbox{\offinterlineskip
\halign{&\vrule#&\strut\enspace\hfil#\enspace\cr
\omit&\omit&\omit&$NN$&\omit&$NF$&\omit&$FN$&\omit&$FF$&\omit\cr
\omit&\omit&\multispan9\hrulefill\cr
\omit&\omit&height2pt&\omit&&\omit&&\omit&&\omit&\cr
\omit&$N$&&$-1$&&1&&1&&$-1$&\cr
\omit&\omit&height2pt&\omit&&\omit&&\omit&&\omit&\cr
\omit&\omit&\multispan9\hrulefill\cr
\omit&\omit&height2pt&\omit&&\omit&&\omit&&\omit&\cr
\omit&$F$&&1&&$-1$&&$-1$&&1&\cr
\omit&\omit&height2pt&\omit&&\omit&&\omit&&\omit&\cr
\omit&\omit&\multispan9\hrulefill\cr }}
$$
where the rows and columns are labelled by Picard's and Q's {\it
pure strategies}\/ (moves), respectively; $F$ denotes a flip and $N$
denotes no flip; and the numbers in the matrix are Picard's
payoffs:  1 indicating a win and $-1$ a loss of a one currency
unit. Q's payoffs can be obtained by reversing the signs in the
above matrix (this is the defining feature of a  zero sum game).

Example:  Q's strategy is to flip the spin on his first turn and
then not flip it on his second, while Picard's strategy is to not
flip the spin on his turn.  The result is that the state of the
spin is, successively: $U$, $D$, $D$, $D$, so Picard wins.

 It is natural to define a two dimensional vector space $V$
with basis $(U,D)$ and to represent players' strategies by
sequences of $2\negthinspace\times\negthinspace 2$ matrices.  That
is, the matrices
$$
F := \bordermatrix{
                       &\scriptstyle{U} & \scriptstyle{D}          \cr
       \scriptstyle{U} &       0        &        1                 \cr
       \scriptstyle{D} &       1        &        0                 \cr
                  }
\qquad\hbox{and}\qquad N := \bordermatrix{
                       &\scriptstyle{U} & \scriptstyle{D}          \cr
       \scriptstyle{U} &       1        &        0                 \cr
       \scriptstyle{D} &       0        &        1                 \cr
                  }
$$
correspond to flipping and not flipping the spin, respectively,
since we define them to act by left multiplication on the vector
representing the state of the spin. A general {\it mixed strategy}\/
consists in a  linear combination of $F$ and $N$, which acts as a
$2\negthinspace\times\negthinspace2$  matrix:
$$
\bordermatrix{
                       &\scriptstyle{U} & \scriptstyle{D}          \cr
       \scriptstyle{U} &      1-p       &        p                 \cr
       \scriptstyle{D} &       p        &       1-p                \cr
                  }
$$
if the player flips the spin with probability $p \negthinspace\in\negthinspace [0,1]$. A
sequence of mixed actions puts the state of the electron into a
convex linear combination $a\, U + (1\negthinspace-\negthinspace a)\,
D$, $0 \le a \le 1$, which
means that if the spin is measured the electron will be in the
spin--up state with probability $a$\/. Q\,,  having studied quantum
theory, is utilizing a { \it quantum strategy}\/, implemented as a
sequence of unitary, rather than stochastic, matrices. In standard
Dirac notation  the basis of $V$ is written $(|U\rangle,|D\rangle
) $. A {\it pure\/} quantum state for the electron is a linear
combination $a\,|U\rangle + b\,|D\rangle,\,a,\,b \in \mathbb{C}$,
$a\,\overline{a} + b\,\overline{b} = 1$, which means that if the spin
is measured, the electron will be in the spin--up state with
probability $a\,\overline{a}$. Since the electron starts in the
state $|U\rangle$, this is the state of the electron if Q's first
action is the unitary operation
$$
U_1 = U(a,b) := \bordermatrix{
                       &\scriptstyle{U} & \scriptstyle{D}          \cr
       \scriptstyle{U} &       a        &     \phantom{-}b                 \cr
       \scriptstyle{D} &  \overline{b}  &  -\overline{a}           \cr
             }.
$$
Captain Picard is  utilizing a {\it classical mixed
  strategy}\/ (probabilistic) in which he flips the spin with
probability $p$\, (has he  preferred drill to studying quantum
theory?). After his action the electron is in a { mixed\/} quantum
state, i.e., it is in the pure state $b\,|U\rangle + a\,|D\rangle$
with probability $p$\/ and in the pure state $a\,|U\rangle +
b\,|D\rangle$ with probability $1\negthinspace-\negthinspace p$\,.
Mixed states are
conveniently represented as {\it density matrices\/}, elements of
$V\negthinspace\otimes \negthinspace V^{\dagger}$ with trace 1;
the diagonal entry $(k,k)$ is the probability that
the system is observed to be in the state $|\psi_k\rangle$.  The
density matrix for a pure state $|\psi\rangle\negthinspace
\in\negthinspace V$ is the projection matrix
$|\psi\rangle\langle\psi|$ and the density matrix for a mixed
state is the corresponding convex linear combination of pure
density matrices. Unitary transformations act on density matrices
by conjugation: the electron starts in the pure state $\rho_0 =
|U\rangle \langle U|$
 and Q's first action puts it into the pure state:
$$
\rho_1 = U_1^{\vphantom\dagger} \rho_0\, U_1^{\dagger}
       = \left( \begin{array}{cc} a\,\overline{a} & a\,\overline{b} \\
                   b\,\overline{a} & b\,\overline{b} \end{array}\right).
$$
Picard's mixed action acts on this density matrix, not as a
stochastic matrix on a probabilistic state, but as a convex linear
combination of unitary (deterministic) transformations:
$$
\rho_2 = p\, F \rho_1 F^{\dagger} + (1\negthinspace -\negthinspace p)\, N \rho_1 N^{\dagger}=
$$
$$\left(
        \begin{array}{cc} p\,b\,\overline{b} + (1\negthinspace-\negthinspace p)\,a\,\overline{a} &
                   p\,b\,\overline{a} + (1\negthinspace-\negthinspace p)\,a\,\overline{b} \\
                   p\,a\,\overline{b} + (1\negthinspace-\negthinspace p)\,b\,\overline{a} &
                   p\,a\,\overline{a} + (1\negthinspace-\negthinspace p)\,b\,\overline{b} \end{array}
                 \right).
$$
For $p\negthinspace =\negthinspace \frac{1}{2}$\, the diagonal
elements of $\rho_2$ are equal to $\frac{1}{2}$\,.  If the game were
to end here, Picard's strategy would ensure him the expected
payoff of 0, independently of Q's strategy. In fact, if Q were to
employ any strategy for which $a\,\overline{a} \ne
b\,\overline{b}$, Picard could obtain the expected payoff of
$|a\,\overline{a} - b\,\overline{b}| > 0$ by setting
$p = 0,1$
according to whether $b\,\overline{b} > a\,\overline{a}$, or the
reverse. Similarly, if Picard were to choose $p\ne\negthinspace
\frac{1}{2}$\,, Q could obtain the expected payoff of $|2p - 1|$ by
setting $a\negthinspace =\negthinspace 1$ or $b\negthinspace
=\negthinspace 1$ according to whether $p\negthinspace <
\negthinspace\frac{1}{2}$\,, or the reverse. Thus the mixed/quantum
equilibria for the two--move game are pairs $\bigl([\frac{1}{2}\,F +
\frac{1}{2}\,N],[U(a,b)]\bigr)$ for which $a\,\overline{a}
= b\,\overline{b}=\frac{1}{2} $\,
and the outcome is the same as if
both players utilize optimal mixed strategies. {But Q has another
move at his disposal ($U_3$) which again transforms the state of
the electron by conjugation to $\rho_3 =
U_3^{\vphantom\dagger}\,\rho_2\, U_3^{\dagger}$. If Q's strategy
consists of $U_1 = U(1/\sqrt{2},1/\sqrt{2}) = U_3$, his first
action puts the electron into a simultaneous eigenstate of both
$F$ and $N$ (eigenvalue 1), which is therefore invariant under
 any mixed strategy $p\,F + (1\negthinspace-\negthinspace p)N$ of Picard. His second
move inverts his first move and produces  $\rho_3 =
|U\rangle\langle U|$. That is, with probability 1 the electron
spin is up!  Since Q can do no better than to win with probability
1, this is an optimal quantum strategy for him. All the pairs
$$\bigl([p\,F + (1\negthinspace-\negthinspace p)N],
       [U(1/\sqrt{2},1/\sqrt{2}),U(1/\sqrt{2},1/\sqrt{2})]\bigr)$$
are mixed/quantum equilibria, with value $-1$ to Picard; this is
why he loses every game. We think that this hypothetical story
convinces the reader that quantum games should be studied
thoroughly in order  to prevent analogous events from  shaping
his/her destiny let alone other aspects. The practical lesson that
the above example teaches is that quantum theory may offer
strategies that at least in some cases give advantage over
classical strategies. Therefore physicist and game theorists
should find answers to the following five questions.
\begin{itemize}
\item Is the idea of quantum game feasible?
\item Under what conditions some players may be able to take the
advantage of quantum phenomena?
\item Are there genuine quantum games that have no classical
counterparts or origin?
\item Are protocols for playing quantum game against human player secure
against cheating?
\item Can the formalism be generalized to include other non--Boolean logic based systems?
\end{itemize}
Finding answers to the above questions is challenging and
intriguing. It is anticipated that answers to these questions will
have a profound impact on the development of quantum theory,
quantum information processing and technology. Unfortunately, at
this stage it we are not able to give any definite answer and, in
fact, we have no idea in what direction we should look to find
them. Nevertheless one can present some strong arguments for
developing quantum theory of games. Modern technologies are
developed mostly due to investigation into the quantum nature of
matter. The results of recent experiments in nanotechnology,
quantum dots and molecular physics are very promising. This means
that we sooner or later may face situations analogous to captain
Picard's if we are not on alert.   Many cryptographic and
information processing problems can be reformulated in game--like
setting. Therefore quantum information and quantum cryptography
should provide us with cases in point. It is obvious that some
classical games can be implemented in such a way that the set of
possible strategies would include strategies that certainly
deserve the the adjective quantum \cite{Du1, Pie}. Such games can
certainly be played in a  laboratory. This process is often
referred to as quantization of the respective standard game. But
this is an abuse of language: we are in fact defining a new game.
In the classical setting the problems of security and honesty are
usually well defined. Realistic quantum cryptography systems and
quantum networks (BBN, Harvard and Boston Universities are already
building the DARPA Quantum Network \cite{Chip}) will certainly
provide us with examples of genuine quantum games and strategies.
Unlike, in quantum game theory the problem is much more involved.
In many cases it can be settled in the "classical way"
(e.g\mbox{.} by selecting arbiters or sort of clearinghouses)  but
if you admit quantum strategies in less definite setting of
actually being developed technologies it may be even difficult to
name dishonesty. If quantum games should ever be applied outside
physical laboratories a lot of technical problems must be solved.
Security of quantum games is only one of them but it already
involves error corrections, quantum state tomography and methods
of communications and preparations of quantum systems forming the
"quantum board" and the necessary "quantum memory". We envisage
that critical analysis of already proposed  quantum information
processing protocols must be done to this end. One of the main
objectives would be a definition of (possibly universal)
primitives necessary for realistic quantum games. Quantum
phenomena probably play important role in biological and other
complex systems and, although this point of view is not commonly
accepted, quantum games may turn out to be an important tool for
the analysis of various complex systems. Genetic algorithms and
DNA computation can also be used to implement games and quantum
games may be the most promising  field \cite{Wood}. Massive
parallel DNA processing would allow to play simultaneously
trillions of games. Noncommutative propositions are
characteristic of various situations not necessary associated with
quantum systems. In fact, the richness of possible structures is
immense. There are suggestions that quantum--like description of
market phenomena may be more accurate than the classical
(probabilistic) one \cite{Wai}. The quantum morphogenesis
\cite{Aer} shows one possible way of generalization of the
formalism that may find application in social sciences.

\section{Quantum game theory}

Basically, any quantum system that can be manipulated  by at least
one party and where the utility of the moves can be reasonably
defined, quantified and ordered may be conceived as a quantum
game. The quantum system may  be referred to as a {\it quantum
board}\/ although the term {\it universum of the game}\/ seems to
be more appropriate \cite{Bug}. We will suppose that all players
know the state of the game at the beginning and  at some crucial
stages that may depend an the game being played. This is a subtle
point because it is not always possible to identify the state of a
quantum system let alone the technical problems with actual
identification of the state (one can easily give examples of
systems that are only partially accessible to some players
\cite{Bug2}). A "realistic" quantum game should include measuring
apparatuses or information channels that provide information on
the state of the game at crucial stages and specify the way of its
termination. We will neglect these nontrivial issues here.
Therefore we will suppose that a {\it two--player quantum game}\/
$\Gamma\negthinspace =\negthinspace({\cal
H},\rho,S_A,S_B,P_A,P_B)$ is completely specified by the
underlying Hilbert space ${\cal H}$ of the physical system, the
initial state $\rho\negthinspace\in\negthinspace {\cal S}({\cal
H})$, where ${\cal S}({\cal H})$ is the associated state space,
the sets $S_A$ and $S_B$ of permissible quantum operations of the
two players, and the { pay--off (utility) functions}\/ $P_A$ and
$P_{B\/}$, which specify the pay--off for each player. A {\it
quantum strategy}\/ $s_A\negthinspace\in\negthinspace S_A$,
$s_B\negthinspace\in\negthinspace S_B$ is a collection of
admissible quantum operations, that is the mappings of the space
of states onto itself.
One usually supposes that they are  completely
positive trace--preserving maps. The quantum game's definition may
also include certain additional rules, such as the order of the
implementation of the respective quantum strategies or restriction
on the admissible communication channels, methods of stopping the
game etc. We also exclude the alteration of the pay--off during the
game. The generalization for the N players case is obvious.
 Schematically we have:
$$
\rho \mapsto (s_{A},s_{B}) \mapsto  \sigma \Rightarrow (P_{A},
P_{B})\,.$$

The following concepts  will be used in the remainder of this
paper. These definitions are completely analogous to the corresponding
definitions in standard game theory \cite{Osb, Str}. The adjective
quantum gives no extra meaning to them. A strategy $s_A$ is called
a {\it dominant strategy}\/ of Alice if
\begin{eqnarray*}
        P_A(s_A,s_B')
        &\geq&
        P_A(s_A',s_B')
\end{eqnarray*}
for all $s_A'\negthinspace\in\negthinspace S_A$, $s_B'\negthinspace\in\negthinspace
 S_B$. Analogously we can define a
dominant strategy for Bob. A pair $(s_A,s_B)$ is said to be an
{\it equilibrium in dominant strategies}\/ if $s_A$ and $s_B$ are
the players' respective dominant strategies. A combination of
strategies $(s_A,s_B)$ is called a {\it Nash equilibrium}\/ if
\begin{eqnarray*}
        P_A(s_A,s_B)&\geq& P_A(s_A',s_B)\,,\\
        P_B(s_A,s_B)&\geq& P_B(s_A,s_B')\, .
\end{eqnarray*}
A pair of strategies $(s_A, s_B)$ is called {\it Pareto
optimal}\/, if it is not possible to increase one player's pay--off
without lowering the pay--off of the other player. A solution in
dominant strategies is the strongest solution concept for a
non--zero sum game. For example, in the popular Prisoner's Dilemma
game \cite{Osb, Str}:
$$ \begin{array}{c|cc}
     & \mbox{Bob}: C & \mbox{Bob}: D \\
    \hline
    \mbox{Alice}: C & (3,3) & (0,5) \\
    \mbox{Alice}: D & (5,0) & (1,1)
  \end{array}
$$
where the numbers in parentheses represent the row (Alice) and
column (Bob) player's payoffs, respectively. Defection (D) is the
dominant strategy, as it is favorable regardless what strategy the
other party chooses.

In general the optimal strategy depends on the strategy chosen by
the other party. A Nash equilibrium implies that neither player
has a motivation to unilaterally alter his/her strategy from this
kind of equilibrium solution, as this action will lower his/her
pay--off. Given that the other player will stick to the strategy
corresponding to the equilibrium, the best result is achieved by
also playing the equilibrium solution. The concept of Nash
equilibria is therefore of paramount importance to studies of
non--zero--sum games. It is, however, only an acceptable solution
concept if the Nash equilibrium is not unique (this happens very
often). For games with multiple Nash equilibria we have to find a
way to eliminate all but one of  them. Therefore a Nash
equilibrium is not necessarily an efficient and satisfactory one.
We say that an equilibrium is Pareto optimal if there is no other
outcome which would make both players better off. But usually
there are no incentives to adopt the Pareto optimal strategies. In
the Prisoner's Dilemma the Pareto equilibrium is reached if both
players adopt the strategy C, but they are afraid of being
outwitted by the opponent's playing D. If the same game is
repeated many times the situation changes because the players may
communicate by changing strategies and learning is possible. To
this end both players should  adopt mixed strategies. One can
prove that any game has  a Nash equilibrium in the class of mixed
strategies \cite{Osb, Str}.

\section{Quantum games in action: a review recent results}

Quantum game theory attracted much attention since the
prescription for quantization of games has been put forward  by
Eisert, Wilkens and Lewenstein  \cite{Eis}. It was subsequently
generalized by Marinatto and Weber \cite{Mar}. This general
setting was described above. Actually, it   can be applied to any
$2\negthinspace\times\negthinspace n$ games (each player has $n$
strategies and the players' actions are represented by $U(n)$ or
$SU(n)$ operators). This prescription has been used for
"quantizing" various classical games (Prisoner's Dilemma
\cite{Eis,Du2}, The Monty Hall Problem \cite{FA1, Ari}, Battle of
Sexes \cite{Naw, Du3, Du4},  Stag Hunt Game \cite{Toy}, Rock,
Scissors and Paper \cite{IT2, Sto}, Coordination Problem
\cite{Hub}, Duopoly Problem \cite{IT3}, duels \cite{FA2} etc. The
results show that, in general, the "quantization process" and
relations to the background classical problems are not unique.
Nash equilibria can be found but, as in the classical problems, in
most cases they are not Pareto optimal. Lee and Johnson have shown
that playing games quantum mechanically can be more efficient and
giving a saturation of the upper bound on the efficiency
\cite{L-J}. From their work it can be deduced that there are
quantum versions of the minimax theorem for zero sum quantum games
and the Nash equilibrium theorem for general static quantum games.
There are many unexplored connections between quantum information
theory and other scientific models. Quantum game theory offers
tools in analysis of phenomena that usually are not considered as
physical processes. Theory of information can be used for analysis
of algorithms that describe player's strategies and tactics but
classical games form only a small? subclass of games that can be
played in quantum information media. If we ignore technological
problem then we can extend this subclass so that exploration of
quantum phenomena is possible. There are two obvious modifications
of classical simulation games.
\begin{description}
\item[1 -- prequantization:] Redefine the game so that it became a
reversal operation on qubits representing player's strategies.
This allows for quantum coherence of strategies\footnote{This may
result from nonclassical strategies or classically forbidden
measurements of the state of the game}.
\item[2 -- quantization:]
Reduce \label{punktdwa} the number of qubits and allow arbitrary
unitary\footnote{At least one of the performed operations should
not be equivalent to a classical one } transformation so that the
basic feature of the classical game are preserved.
\end{description}
Any game modified in this way is in fact a quantum algorithm that
usually allows for more effective information processing than the
starting game. Actually, any quantum computation is a potential
quantum game if we manage to reinterpret it in game--theoretical
terms. To illustrate the second method let us consider Wiesner's
counterfeit--proof banknote \cite{Wie}. This is the first quantum
secrecy method (elimination of effective eavesdropping). As a
quantum game it consists in a finite series of sub--games
presented in Fig\mbox{.} \ref{goresyt1}. An arbiter Trent produces a pair
of random qubits $|\psi_{T}\hspace{-.1em}\rangle$ and
$|\psi_{T'}\negthinspace\rangle$. The polarization of the qubit
(strategy) $|\psi_{T}\hspace{-.1em}\rangle$ is known to Trent and
is kept secret. The qubit  $|\psi_{T'}\negthinspace\rangle$ is
ancillary.  Alice qubit $|\psi_{\negthinspace A}\rangle$ describes
her strategies $|\text{I}\rangle$ and $|0\rangle$. The first move
is performed by Alice. Her strategy $|\text{I}\rangle$ consists in
switching the Trent's qubits $|\psi_{T}\hspace{-.1em}\rangle$ and
$|\psi_{T'}\negthinspace\rangle$. The strategy $|0\rangle$
consists in leaving the Trent's qubits intact. These moves form
the controlled--swap gate \cite{Nie}. Her opponent Bob wins only
if after the game Trent learns that his qubit
$|\psi_{T}\hspace{-.1em}\rangle$ has not been changed.

\begin{figure}[h]
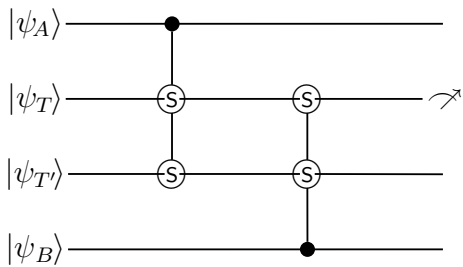

\begin{center}
\phantom{a}\vspace{15ex} \psset{linewidth=.7pt}
\rput(-2.7,3){\rnode{A}{$|\psi_{\negthinspace
A}\rangle$\hspace{1pt}}} \cnode*(-.9,3){.1}{B}
\rput(2.7,3){\rnode{C}{}}
\rput(-2.7,2){\rnode{D}{$|\psi_{T}\hspace{-.1em}\rangle$\hspace{1pt}}}
\rput(-.9,2){\rnode{E}{\circs}} \rput(.9,2){\rnode{F}{\circs}}
\rput(2.7,2){\rnode{G}{\hspace{.15em}\meter}}
\rput(-2.7,1){\rnode{H}{$|\psi_{T'}\negthinspace\rangle$\hspace{1pt}}}
\rput(-.9,1){\rnode{I}{\circs}} \rput(.9,1){\rnode{J}{\circs}}
\rput(2.7,1){\rnode{K}{}}
\rput(-2.7,0){\rnode{L}{$|\psi_{\hspace{.15em}\negthinspace
B}\hspace{-.05em}\rangle$\hspace{1pt}}} \cnode*(.9,0){.1}{M}
\rput(2.7,0){\rnode{N}{}} \ncline[nodesep=0pt]{-}{A}{C}
\ncline[nodesep=0pt]{-}{D}{E} \ncline[nodesep=0pt]{-}{E}{F}
\ncline[nodesep=0pt]{-}{F}{G} \ncline[nodesep=0pt]{-}{H}{I}
\ncline[nodesep=0pt]{-}{I}{J} \ncline[nodesep=0pt]{-}{J}{K}
\ncline[nodesep=0pt]{-}{L}{N} \ncline[nodesep=0pt]{-}{B}{E}
\ncline[nodesep=0pt]{-}{E}{I} \ncline[nodesep=0pt]{-}{F}{J}
\ncline[nodesep=0pt]{-}{J}{M}
\end{center}
\caption{Quantum identification game constructed from two
controlled--swap gates (Wiesner's money).} \label{goresyt1}
\end{figure}
To win Bob must always begin with with a strategy identical to the
one used by Alice. If there is no coordination of moves between
Alice and Bob the probability of Bob's success exponentially
decreases with growing number of sub--games being played and is
negligible even for a small number of sub--games. Alice and Bob's
strategies are classical but due to the prequantization process
eavesdropping is not possible if Trent uses arbitrary
polarizations $|\psi_{T}\hspace{-.1em}\rangle=
|0\rangle+z\,|\text{I}\rangle$, $z\in\overline{\mathcal{C}}\simeq
S_2$ (in the projective nonhomogeneous coordinates). This game can
be quantized by elimination of the ancillary qubit
$|\psi_{T'}\negthinspace\rangle$. Then Alice and Bob strategies
should be equivalent to controlled--Hadamard gates \cite{Nie}. In
this case Trent's qubit is changed only if Alice adopts the
strategy $|\text{I}\rangle$ that result in
$|\psi_{T}\hspace{-.1em}\rangle=
|0\rangle+z\,|\text{I}\rangle\longrightarrow
|0\rangle+\frac{1-z}{1+z}\,|\text{I}\rangle$  (quantum Fourier
transform), see Fig\mbox{.} \ref{goresyt2}. The actual Wiesner's idea was
to encode the secret values of $|\psi_{T}\hspace{-.1em}\rangle$
that result from Alice moves in the series of sub--games on an
otherwise numbered banknote. In addition, the issuer  Trent takes
over the role of Alice and records the values of
$|\psi_{T}\hspace{-.1em}\rangle$ and $|\psi_{\negthinspace
A}\rangle$ with the label being the number of the banknote.  The
authentication of the banknote is equivalent to a success in the
game when Bob's strategy is used against that recorded by Trent
(if Bob wins then his forgery is successful).

\begin{figure}[h]
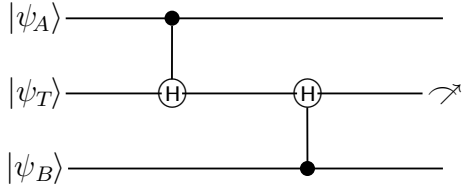

\begin{center}
\phantom{a}\vspace{11ex} \psset{linewidth=.7pt}
\rput(-2.7,2){\rnode{A}{$|\psi_{\negthinspace
A}\rangle$\hspace{1pt}}} \cnode*(-.9,2){.1}{B}
\rput(2.7,2){\rnode{C}{}}
\rput(-2.7,1){\rnode{D}{$|\psi_{T}\hspace{-.1em}\rangle$\hspace{1pt}}}
\rput(-.9,1){\rnode{E}{\circh}} \rput(.9,1){\rnode{F}{\circh}}
\rput(2.7,1){\rnode{G}{\hspace{.15em}\meter}}
\rput(-2.7,0){\rnode{H}{$|\psi_{\hspace{.15em}\negthinspace
B}\hspace{-.05em}\rangle$\hspace{1pt}}} \cnode*(.9,0){.1}{I}
\rput(2.7,0){\rnode{J}{}} \ncline[nodesep=0pt]{-}{A}{C}
\ncline[nodesep=0pt]{-}{D}{E} \ncline[nodesep=0pt]{-}{E}{F}
\ncline[nodesep=0pt]{-}{F}{G} \ncline[nodesep=0pt]{-}{H}{J}
\ncline[nodesep=0pt]{-}{B}{E} \ncline[nodesep=0pt]{-}{F}{I}
\end{center}
\caption{Quantum identification game constructed from two
controlled--Hadamard gates.} \label{goresyt2}
\end{figure}

The introduction of classically impossible strategies results in
better security against quantum attack (pretending to be Alice).
Eavesdropping of the state $|\psi_{T}\hspace{-.1em}\rangle$
modified by Alice's strategy is ineffective  even if Trend limits
himself to polarizations from the set
$\{|0\rangle,|\text{I}\rangle\}$. It is possible that an analogous
reduction of qubits allows to  exponentially reduce the complexity
of quantum algorithms. Therefore quantum games may sometimes be
the only feasible alternatives if the classical problems are
computationally to complex to be ever implemented. Description of
games against Nature is far more complicated. It is not easy to
show that they do not contradict known natural laws or are
actually being played.  For example, let us consider a
prequantized version of the Maxwell's Demon game against Nature
\cite{plenio}. Demon acting in accord with physical laws tries to
build a Szilard's engine. The Demon fails because Nature erases
information and this is an energy consuming process. Such
arguments work also in biology and social sciences. The classical
theory of interacting particles localized at nodes of crystal
lattices results in quantum model of collective phenomenon known
as phonon. Phonons do not exist outside the crystal lattice. May
humans and animals form classical ingredients of large quantum
entities? May Penrose will not be able to find consciousness at
the sub--neuronic microtubular level \cite{Penrose} because it is
localized at the complex multi--neuron level? Do we try to
convince ourselves that there are living actors in TV sets?
\subsection{Quantum games in economics and social sciences}
Modern game theory has its roots in economics and social sciences
and one should not be surprised by number of attempts at
quantizing classical problems. In the "standard" quantum game
theory one tries in some sense to quantize an operational
description of "classical" versions of the game being analyzed. It
usually enlarges  the set admissible strategies in a nontrivial
way. Piotrowski and S\l adkowski follow a different way.
Technological development will sooner or later result in
construction of quantum computers. If one considers the number of
active traders, the intensity of trade contemporary markets and
their role in the civilization  then one must admit that the {\it
the market game}\/ is the biggest one ever played by humans.  How
would look a market cleared by a quantum computer quantum network
or other quantum device? They propose to describe market players
strategies in terms of state vectors $|\psi\rangle$ belonging to
some Hilbert space ${\cal H}$ \cite{PS1, PS2}. The probability
densities of revealing the players, say Alice and Bob, intentions
are described in terms of random variables $p$ and $q$:
\begin{equation*}
\label{eigenstosc} \frac{|\langle q|\psi\rangle_A|^2}{\phantom{}_A
\langle\psi|\psi\rangle_A}\, \frac{|\langle
p|\psi\rangle_B|^2}{\phantom{}_B \langle\psi|\psi\rangle_B}\;d q d
p\, ,
\end{equation*}
where $\langle q|\psi\rangle_A$\/ is the probability amplitude of
offering the price $q$ by Alice who wants to buy and the demand
component of her state is given by
$|\psi\rangle_A\in\mathcal{H}_{A}$\/. Bob's amplitude $\langle
p|\psi\rangle_B$\/ is interpreted in an analogous way (opposite
position). Of course, the "intentions" $q$ and $p$ not always
result in the accomplishment of the transaction \cite{PS1}.
According to  standard risk theory  it seems reasonable to define
the observable of {\it the risk inclination operator}\/:
$$
H(\mathcal{P}_k,\mathcal{Q}_k):=\frac{(\mathcal{P}_k-p_{k0})^2}{2\,m}+
                     \frac{m\,\omega^2(\mathcal{Q}_k-q_{k0})^2}{2}\,,
 $$ \noindent where $p_{k0}:=\frac{
\phantom{}_k\negthinspace\langle\psi|\mathcal{P}_k|\psi\rangle_k }
{\phantom{}_k\negthinspace\langle\psi|\psi\rangle_k}\,$,
$q_{k0}:=\frac{
\phantom{}_k\negthinspace\langle\psi|\mathcal{Q}_k|\psi\rangle_k }
{\phantom{}_k\negthinspace\langle\psi|\psi\rangle_k}\,$,
$\omega:=\frac{2\pi}{\theta}\,$.  $ \theta$ denotes the
characteristic time of transaction \cite{PS3} which is, roughly
speaking, an average time spread between two opposite moves of a
player (e.g.~buying and selling the same asset). The parameter
$m\negthinspace>\negthinspace0$ measures the risk asymmetry between
buying and selling
positions. Analogies with quantum harmonic oscillator allow for
the following characterization of quantum market games. The
constant $h_E$ describes the minimal inclination of the player to
risk. It is equal to the product of the lowest eigenvalue of
$H(\mathcal{P}_k,\mathcal{Q}_k) $ and $2\,\theta$. $2\,\theta $ is in
fact the minimal interval during which it makes sense to measure
the profit.  Except the ground state all the adiabatic strategies
$H(\mathcal{P}_k,\mathcal{Q}_k)|\psi\rangle={const}|\psi\rangle$
are giffens \cite{PS1, Sla} that is goods that do not obey the law
of demand and supply. It should be noted here that in a general
case the operators $\mathcal{Q}_k $ do not commute because traders
observe moves of other players and often act accordingly. One big
bid can influence the market at least in a limited time spread.
Therefore it is natural to apply the formalism of noncommutative
quantum mechanics where one considers
$$ [ x^{j},x^{k}] = \text{i} \,\Theta ^{jk}:=\text{i}\,\Theta \,\epsilon ^{jk}. $$
The analysis of harmonic oscillator in more then one dimensions
 imply that the parameter $\Theta $ modifies the constant
$\hslash_E$ $\rightarrow \sqrt{\hslash_E^{2} + \Theta ^{2}} $ and,
accordingly, the eigenvalues of $H(\mathcal{P}_k,\mathcal{Q}_k)$.
This has the natural interpretation that moves performed by other
players can diminish or increase one's inclination to taking risk.
 When a game allows a
great number of players in it is useful to consider it as a
two--players game: the trader $|\psi\rangle_{k}$ against the Rest
of the World (RW). The concrete algorithm $\mathcal{A}$ may allow
for an effective  strategy of RW (for a sufficiently large number
of players a single player  would not have much influence on the
form of the RW strategy). If one considers the RW strategy it make
sense to declare its simultaneous demand and supply states because
for one player RW is a buyer and for another it is a seller. To
describe such situation it is convenient to use the Wigner
formalism. If the market  continuously measures the same strategy
of the player, say the demand $\langle q|\psi\rangle $, and the
process is repeated sufficiently often for the whole market, then
the prices given by some algorithm  do not result from the
supplying strategy $\langle p|\psi\rangle $ of the player. The
necessary condition for determining the profit of the game is the
transition of the player to the state $\langle p|\psi\rangle $.
If, simultaneously, many of the players change their strategies
then the quotation process may collapse due to the lack of
opposite moves. In this way the quantum Zeno effects explains
stock exchange crashes. Another example of the quantum market Zeno
effect is the
stabilization of prices of an asset provided by a monopolist. \\
But one does not need any sophisticated equipment or technology to
apply quantum theory and quantum games in economics and social
sciences. In fact, Lambertini claims that quantum mechanics and
mathematical economics are isomorphic \cite{Lam}. Therefore one
should expect that various quantum tools as the quantum
morphogenesis \cite{Aer} would be invented and used to describe
social phenomena. An interesting analysis was done by Arfi who
proposes to use quantum game for wide spectrum of problems in
political sciences \cite{Arf}. Quantum game theory may help
solving some philosophical paradoxes, c.f\mbox{.} the quantum
solution to the notorious Newcomb's paradox (free will  dilemma)
\cite{PS4}.\\

 Strategies adopted by social groups or their
individual members usually seem to be unspeakable and elusive.
Efforts to imitate them often fail. Impossible to clone quantum
strategies have analogous properties \cite{Nie}. On the other
side, customs, habits and memories are so durable that
possibilities of effacing them are illusory. This resembles the no
deleting theorem for quantum states (strategies) \cite{Pati}. Both
theorems describe the same forbidden process expressed in reverse
chronological order
$|\psi\rangle|0\rangle\nleftrightarrow|\psi\rangle|\psi\rangle$.
An interesting thermodynamical discussion of this impossibility is
given in \cite{horo}. These analogies cannot be explained in the
classical paradigm.

\subsection{Quantum games in biology}
Living organism may in fact behave in quantum--like way. This  may
be caused by at least two factors. First, quantum entanglement and
decoherence  may affect various molecular processes. Second,
quantum--like description of dynamics in a population of
interacting individuals may be more accurate than the
probabilistic one. Therefore a cautious speculation on the
possibility that the natural world might already be exploiting the
advantages of quantum games on the macroscopic scale may be in
place. Maynard Smith in his book \textit{Evolution and the Theory
of Games} \cite{smith} discusses an evolutionary approach in
classical game theory. The concept of evolutionary stability
stimulated the development of evolutionary game theory. Iqbal and
Toor showed in a series of papers \cite{IT1,IT2, IT4,IT5,IT6} that
the presence of entanglement, in asymmetric as well as symmetric
bimatrix games, can disturb the evolutionary stability expressed
by the idea of evolutionary stable strategies. Therefore
evolutionary stability of a symmetric Nash equilibrium can be made
to appear or disappear by controlling entanglement in symmetric
and asymmetric bimatrix games. It shows that the presence of
quantum mechanical effects may have a deciding role on the
outcomes of evolutionary dynamics in a population of interacting
entities. They suggest that a relevance of their ideas may be
found in the studies of the evolution of genetic code at the dawn
of life and evolutionary algorithms where interactions between
individual of a population may be governed by quantum effects. The
nature of these quantum effects, influencing the course of
evolution, will also determine the evolutionary outcome. Therefore
Darwin's idea of natural selection may be relevant even for
quantum systems. Another class of problem concerns replication in
biology \cite{Kau}.  The role of DNA and its replication still
waits for explanation.  Game theory and quantum game theory offer
interesting and powerful tools to this end the results will
probably find their applications in computation, physics, complex
system analysis and cognition sciences \cite{Pat1, Pat2, Home}.
Neuroeconomics \cite{Gli1, Gli2}, a discipline that aims at
proving that economic theory may provide an alternative to the
classical Cartesian model of the brain and behavior, is a source
of fascinating topics for a debate.  Is there a {\it quantum
neuroeconomics}\/ \cite{PS5}? We may expect a rich dialogue
between theoretical neurobiology and quantum logic \cite{McC}.
\subsection{Quantum games and information processing}
Quantum theory of information is certainly a serious challenge to
the standard game theory  and will probably stimulate the research
in quantum game theory. Most of the cryptographic problems are in
fact games, sometimes in camouflage.  Analysis and design of
cryptographic primitives can  also be perceived as games so their
quantum counterparts are quantum games, e.g\mbox{.} quantum key
distribution, quantum coin tossing \cite{Keyl,Chip} or the
coordination problem in distributive computing \cite{Fit}. Coin
tossing protocols form an important class of cryptographic
primitives. They are used to define a random bit among  separated
parties. Classical coin flipping can implemented by a trusted
arbiter or by assuming that the players have limited computational
power. If the players have unlimited computational power then no
classical coin flipping protocol is possible because any such
protocol represents a two player game and, according to game
theory, there always is a player with a winning strategy. By
contrast, in a quantum world the existence of coin flipping
protocols is not ruled out even  by unlimited computational power.
This is because any attempt by a player to deviate (cheat) from
the protocol can disturb the quantum states, and therefore be
detected by the adversary. But this is far from being the whole
story. There are quantum games that live across the border of our
present knowledge. For example, consider some classical or quantum
problem $X$. Let us define the game $kXcl$: you win if and only if
you solve the problem (perform the task) $X$ given access to only
$k$ bits of information. The quantum counterpart reads: solve the
problem $X$ on a quantum computer or other quantum device given
access to only $k$ bits of information. Let us call the game
$kXcl$ or $kXq$ interesting if the corresponding limited
information--tasks are feasible. Let $OkhamXcl$ ($OkhamXq$)
denotes the minimal $k$ interesting game in the class $kXcl$
($kXq$). Authors of the paper \cite{PS5} described the game played
by a market trader who gains the profit $P$ for each bit (qubit)
of information about her strategy. If we denote this game by $MP$
then
$OkhamM\frac{1}{2}\,cl\negthinspace=\negthinspace2M\frac{1}{2}\,cl$
and for $P\negthinspace>\negthinspace\frac{1}{2}$ the game does
not exist $OkhamMPcl$. They also considered the more effective
game $1M\frac{2+\sqrt{2}}{4}\,q$ for which
$OkhamM\frac{2+\sqrt{2}}{4}\,q\neq1M\frac{2+\sqrt{2}}{4}\,q$ if
the trader can operate on more then one market. This happens
because there are entangled strategies that are more profitable
\cite{EWP}. There are a lot of intriguing questions that can be
ask, for example for which $X$ the meta--game $Okham(OkhamXq)cl$
can be solved or when, if at all,  the meta--problem
$Okham(OkhamXq)q$ is well defined problem. Such problems arise in
quantum memory analysis \cite{Koe}. \\ Algorithmic combinatorial
games, except for cellular automata,  have been completely ignored
by quantum physicists. This is astonishing because at least some
of the important intractable problems might be attacked and solved
on a quantum computer (even such a simple one player game as
Minesweeper in NP--complete \cite{saper}).

\subsection{Quantum games, complexity theory and decision theory}
 What form does the decision theory take
for a quantum player? Almost all quantum acts involve preparing a
system, measuring it, and then receiving some reward (in a more or
less general sense) which is dependent on the outcome of the
measurement. Therefore it should not be astonishing that
game--theoretical analysis of quantum phenomena has far reaching
consequences. Deutsch claims to have derived the Born from
decision--theoretic assumptions  \cite{Deu1, Wall}. In fact he have
defined a quantum game and quantum--mechanical version of decision
theory. What is striking about the Deutsch game  is that rational
agents are so strongly constrained in their behavior that not only
must they assign probabilities to uncertain events, they must
assign precisely those probabilities given by the Born rule. His
proof must be understood in the explicit context of the Everett
interpretation, and that in this context it is acceptable
\cite{Wall}. \\
Roughly speaking, one of the main goals of complexity theory is to
present lower bounds on various resources needed to solve a
certain computational problem. From a cryptographic viewpoint, the
most demanding problem is to prove nontrivial lower bounds on the
complexity of breaking  concrete cryptographic systems. Query
complexity on the other side, is an abstract scenario which can be
thought of as a game. The goal is to determine some information by
asking as few questions as possible see e.g\mbox{.} quantum oracle and
their interrogations \cite{Kash}. A weak form of quantum
interactive proof systems known as quantum Merlin--Arthur games
\cite{Kob, Knil, Watr} defines a whole class of quantum games with
wide application in quantum complexity theory and cryptography.
Here, powerful Merlin presents a proof and Arthur, who is the
verifier, verifies its correctness. The task is to prove a
statement without yielding
anything beyond its validity ({\it zero knowledge proofs}\/).\\

There are games in which the agents' strategies do not have
adequate descriptions in terms of some Boolean algebra of logic
and theory of probability. They can be analyzed according to the
rules of quantum theory and the result are promising, see e.g\mbox{.} the
Wise Alice game proposed in \cite{Grib1, Grib2}. This game is a
simplified version of the quantum bargaining game \cite{PS6}
 restricted to the "quantum board" of the form
$[buy,sell]\negthinspace\times\negthinspace [bid, accept]$.
Quantum semantic games also belong to this class \cite{Piet}.

\subsection{Quantum gambling}
 At the present stage of our technological development it already
is feasible to open {\it quantum casinos}, where gambling at
quantum games would be possible. Of course, such an enterprise
would be costly but if you recall the amount of money spent on
advertising various products it seems to us that it is a worthy
cause. Goldenberg, Vaidman and Wiesner described the following
game based on the coin tossing protocol \cite{Gol}. Alice has two
boxes, $A$ and $B$, which can store a particle. The quantum states
of the particle in the boxes are denoted by $|a\rangle$ and
$|b\rangle$, respectively. Alice prepares the particle in some
state and sends box $B$ to Bob.\\
Bob wins in one of the two cases:
\begin{enumerate}
\item
If he finds the particle in box $B$, then Alice pays him $1$
monetary unit (after checking that box $A$ is empty).
\item
If he asks Alice to send him box $A$ for verification and he finds
that she initially prepared a state  different from
$
|\psi_{0}\rangle = 1/\sqrt{2} \: (|a\rangle + |b\rangle) ,
$
then Alice pays him $R$ monetary units.
\end{enumerate}
In any other case Alice wins, and Bob pays her $1$ monetary unit.
They have analyzed the security of the scheme, possible methods of
cheating and calculated the average gain of each party as a result
of her/his specific strategy. The analysis shows that the protocol
allows two remote parties to play a gambling game, such that in a
certain limit it becomes a fair game. No unconditionally secure
classical method is known to accomplish this task. This game was
implemented by Yong--Sheng Zhang et al, \cite{Zha}. Other
proposals based on properties of non--orthogonal states were put
forward by Hwang, Ahn, and Hwang \cite{Hwa1} and  Hwang and
Matsumoto \cite{Hwa2}. Witte proposed a quantum version of the
Heads or Tails game \cite{Wit}. Piotrowski and S\l adkowski
suggested that although  sophisticated technologies to put  a
quantum market in motion  are not yet available, simulation of
quantum markets and auctions can be performed in an analogous way
to precision physical measurements during which classical
apparatuses are used to explore  quantum phenomena. People seeking
after excitement would certainly not miss the opportunity to
perfect their skills at using "quantum strategies". To this end an
automatic game "Quantum Market" will be sufficient and such a
device can be built up  due to the recent advances in technology
\cite{PS5}. Segre published an interesting detailed analysis of
quantum casinos and a Mathematica packages for simulating quantum
gambling \cite{Seg}. His and others considerations show that
quantum gambling is closely related to quantum logic, decision
theory and can be used for defining a Bayesian theory of quantum
probability \cite{Pit}.

\section{Conclusions}
Games, used  both for entertainment  and scientific aims, are
traditionally modelled as mathematical objects, and therefore are
traditionally seen  as mathematical disciplines. However, all
processes in the real world are physical phenomena and as such
involve noise, various uncertainty factors and, what concerned us
here, quantum phenomena. This fact can be used both for the
benefit and detriment. Works of Deutsch, Penrose and others seem
to be harbingers of the dawn of a quantum game era when consistent
quantum information description would be used not only in physics
and natural sciences but also in social sciences and economics.
The heterogeneity and fruitfulness of quantum computations will
certainly stimulate such interdisciplinary studies and
technological development. Quantum  games broaden our horizons and
offer new opportunities for the technology and economics.  If
human decisions can be traced to microscopic quantum events one
would expect that nature would have taken advantage of quantum
computation in evolving complex brains. In that sense one could
indeed say that quantum computers are playing according to quantum
rules. David Deutsch has proposed an interesting unification of
theories of information, evolution and quanta \cite{Deu2}.
Lambertini put forward arguments for observing Schroedinger cat
like objects on real markets \cite{Lam}. But why quantum social
sciences should emerge just now \cite{Men}? They could have not
emerged earlier because a tournament quantum computer versus
classical one is not possible without technological development
necessary for a construction of quantum computers. Now it seems to
be feasible. Quantum--like approach to market description might
turn out to be an important theoretical tool for investigation of
computability problems in economics or game theory even  if never
implemented in real market \cite{Vel, Wai}. It is tempting to
"quantize"  Karl Popper's ideas \cite{pop} expressed in terms of
language--games \cite{Hint}. Such a revision would determine
regions of quantum falsification of scientific theories
(q--falsification). Should theories that have high falsification
and low q--falsification be regarded as restraining development?
If this is the case then {\em the quantum information processing
paradigm}\/ (see e.g\mbox{.} \cite{horo1})  should replace the
alternative platonism or mysticism \cite{Penrose}.
 Of course, as any other disciplines,
quantum game theory also has its negative sides but there is no
doubt that it will be a crucial discipline for the emerging
information society.
\\\\
{\bf Acknowledgements} We are greatly indebted to prof\mbox{.}
Zbigniew Hasiewicz for critical reading of the manuscript and
helpful remarks.

\end{document}